\documentstyle[prb,floats,aps,amssymb,preprint,epsf,rotate]{revtex} 
\begin{document} 

\title{{\Large  {A numerical study of the development of bulk scale-free
structures upon growth of self-affine aggregates.}}} 
\author{Federico Rom\'a$^{1}$, Claudio M. Horowitz$^{2}$ and Ezequiel V.
Albano$^{2}$.}
\address{1- Departamento de F\'{\i}sica, UNSL, Chacabuco 917, (5700), San Luis, Argentina. }   \address{2-
Instituto de Investigaciones Fisicoqu\'{\i}micas  Te\'oricas y Aplicadas
(INIFTA), UNLP, CONICET,  Casilla de  Correo 16, Sucursal 4, 
(1900) La Plata, Argentina} 
\maketitle

\vskip -1.0 true cm

\begin{abstract}
\vskip -1.0 true cm

During the last decade,
self-affine geometrical properties of many
growing aggregates, originated in a wide variety of processes,
have been well characterized.  However,
little progress has been achieved in the search of a unified description
of the underlying dynamics. Extensive numerical evidence has been given
showing that the bulk of aggregates formed upon ballistic aggregation and
random deposition with surface relaxation processes can be broken down into a
set of infinite scale invariant structures called "trees". These two 
types of aggregates
have been selected because it has been established that they belong
to different universality classes: those of 
Kardar-Parisi-Zhang and Edward-Wilkinson,
respectively. Exponents describing the spatial and temporal scale invariance
of the trees can be related to the classical exponents describing the
self-affine nature of the growing interface. Furthermore, those
exponents allows us to distinguish either the compact or non-compact nature
of the growing trees. Therefore, the measurement of the statistic of the
process of growing trees may become a useful experimental technique for the
evaluation of the self-affine properties of some aggregates.     
\end{abstract} 

\vskip 0.5 true cm 
PACS numbers: 64.60.Ht, 05.40.+j, 61.43.Hv, 05.70.Ln

\newpage  

{\bf I. Introduction.}\\ 

The study of growth processes has recently attracted considerable attention
as a consequence of many factors such as the formulation of the
theory of fractals since the pioneering work of Mandelbrot \cite{1},
the development of new experimental techniques (e.g. scanning
tunneling microscopy) and the  
availability of computer facilities for graphical and
numerical simulations. So far, it is well
established that spatial scaling structures originated from growth
processes are extremely common in nature. In fact, a great variety
of systems exhibit self-affinities over extended ranges of spatial
and temporal scales \cite{3,4,5,6}. Therefore, extensive theoretical and 
experimental research has focused on the characterization of 
the self-affine nature of the structures resulting from 
growth processes \cite{3,4,5,6}.

The phenomenological scaling approach to the dynamic evolution of a
self-affine interface early developed by Family and Vicsek \cite{fam,vics}
has become a useful tool to characterize self-affine roughness.
Considering a flat, $d-1$ dimensional surface at time $t=0$ and pointing
the attention to the growing process that occurs essentially parallel to the
surface, it is possible to assume without loss of generality that there
exists a well defined growth direction and that the interface can be described
by a function $h({\bf x},t)$ that gives the height 
of the interface at time $t$ and position ${\bf x}$. Of course, 
such height is measured from the initial flat
surface at t=0. If the interface cannot be described by a single valued
function of ${\bf x}$, the function $h({\bf x},t)$ gives 
the maximum height of the
interface at ${\bf x}$. Considering a section of the 
sample having a typical size 
$L$ (in each of the $d-1$ dimensions of the surface) the average height of
the interface at time $t$ is defined as:

\begin{equation}
<h(t)>= \frac{1}{(L^{d-1})}  \sum_{{\bf x}} h({\bf x},t) ,  \label{Eq.1}
\end{equation}

\noindent where the summation runs over all ${\bf x}$'s. The interface 
width $w(L,t)$ at time $t$ may be defined by the rms of the 
height fluctuations given by:

\begin{equation}
w(L,t)=\left( \left\langle h^{2}(t)\right\rangle -\left\langle h(t)\right\rangle
^{2}\right) ^{1/2} . \label{Eq.2}
\end{equation}

The Family-Viscek scaling approach assumes that:
\begin{equation}
w(L,t)=L^{\alpha }F(t/L^{z}) , \label{Eq.3}
\end{equation}

\noindent where $F(y)\propto y^{\beta }$ for $y<<1$ 
and $F(y)\rightarrow {\it constant}$ for 
$y>>1$, with $z=\alpha /\beta $. Also $\alpha$, $\beta$ and $z$ are the
roughness, growing and dynamic exponents, respectively.

The dynamic exponent $z$ describes the evolution of a correlated region with
time: initially different parts of the interface are independent, but regions
of correlated roughness form over time and their size grows as $\xi
\propto t^{1/z}$. Thus, for a finite sample of side $L$ and $t\rightarrow \infty $ the width
of the growing interface reaches a statistically stationary state so that $%
w(L)\propto L^{\alpha }$. Furthermore, the overall width of the interface
grows as $t^{\beta }$ until it saturates at $L^{\alpha }$.

In contrast to the progress achieved in the characterization of 
the self-affine behavior of interfaces, little attention has been
drawn to the description of the internal structure of the growing
system. In this work, extensive numerical evidence is
presented showing that the bulk of two archetypical growth models, namely
ballistic deposition and random deposition with surface relaxation, can
effectively be rationalized on the basis of a treeing process; i.e, any
growing structure can be thought as the superposition of individual trees.
These individual trees can be defined as follows: in the case of the ballistic
deposition model (figure 1(a)), a newly deposited particle is assumed
to belong to the same tree as that of the nearest neighbor particle where it is
attached. If the deposited particle has more than one nearest neighbor
belonging to different trees such as particle A in figure 1(a), one of them is
selected at random and the particle is incorporated into that tree . In the
case of the random deposition with surface relaxation model (figure 1(b)), a
newly deposited particle is assumed to belong to the tree corresponding to the
impingement site, before its eventual relaxation. See also figure 1 caption
for further details. 

Those trees that spread out incorporating additional 
growing centers; e.g. capturing particles, developing new branches, 
etc., are said to be "alive" (for example trees 1,2 and 4 in figures
1(a) and 1(b)). In contrast, some other trees may stop their growth due to
shadowing by surrounding growing trees, so they become "dead trees" such as
tree 3 in figures 1(a) and 1(b).  The structure of dead trees is frozen in the
sense that it cannot be modified by any further growth. In order to determine
the distribution of dead trees one starts the growing process following the
specified set of rules and after some fixed time all growing points, i.e. all
sites  of the aggregate where further growth is still possible, are identified
as seeds for these trees. Therefore, the number of such seeds becomes the
initial number of trees. During the subsequent process  the competition
between growing trees dominates the dynamics. Some trees become dead (frozen)
and  eventually only a single tree may remain.

Within this context, the aim of this work is to perform an 
extensive numerical investigation of the dynamics of evolving trees of 
two growing processes, namely Random Deposition with Surface Relaxation
(RDSR) and Ballistic Deposition (BD). Furthermore, tests of the scaling
relationships relating critical exponents of the treeing process to those
describing the self-affine nature of the aggregate are performed
in dimensions $1 \leq d \leq 5$. The RDSR and BD processes have been
selected because they belong to different universality classes, namely the
Edward-Wilkinson (EW) \cite{5,edwi} and  Kardar-Parisi-Zhang (KPZ)
\cite{5,KPZ}, respectively.    

The manuscript is organized as follows: in section II detailed 
definitions and a discussion of the scaling exponents describing the 
treeing process and their relationships to exponents related to the
self-affine nature of the aggregate are reviewed. In section III
a brief description of both the RDSR and BD models and the numerical
simulation technique is provided. The results are presented and 
discussed in section IV while the conclusions are stated in section V.\\

{\bf II. Treeing and self-affinity. Scaling relationships.}\\ 

Let us consider an aggregate that is growing above a \( d \)-dimensional
substrate. Such aggregate is formed by trees that compete with each other
leading to the entire pattern. The structural properties of both the trees and
the resulting entire aggregate are determined by the growth mechanism. In
these context, Racz and Vicsek \cite{RV} have shown that for self-similar
fractals the tree size distribution that results from the competitive growth
process is related to the structure of both the entire aggregate and the
individual trees. Subsequently, this concept was extended to the case of
self-affine objects \cite{MM}.

In the following paragraphs we will briefly outline the well known \cite{RV,MM}
relevant definitions and the scaling relationship in order to establish the
framework for the subsequent numerical study. Pointing our attention to dead
trees of size \( s \) (\( s \) is the number of particles belonging to the
tree) one has that both the rms height (\( h_{s} \)) and the rms
width (\( w_{s} \)) of the trees obey simple power 
laws given by 

\begin{equation}  
h_{s}\propto s^{\nu _{\parallel }} ,  \label{Eq.4}  
\end{equation}

and
\begin{equation}
w_{s}\propto s^{\nu _{\perp }},  \label{Eq.5}
\end{equation}
 
\noindent where \( \nu _{\parallel } \) and \( \nu _{\perp } \) are the
correlation length exponents parallel and perpendicular to the growing
direction of the aggregate \cite{MM}. These exponents may be 
different for self-affine (anisotropic) aggregates, 
while for
self-similar objects \( \nu _{\parallel } \)=\( \nu _{\perp } \).

Assuming that \( N_{T} \) is the total number of particles of the aggregate,
the average particle number \( N \) per unit area of the substrate is given
by
\begin{equation}
N=N_{T}/L^{d},  \label{Eq.6}
\end{equation}

\noindent where particles grow on a \(d \)-dimensional hypercubic
substrate of size L in a $D$-dimensional space.

Defining the number of trees with \( s \) particles \( N \)\( _{s}(N) \)
in the whole aggregate one has that

\begin{equation}
N_{T}=\sum _{s}sN_{s}(N),  \label{Eq.7}
\end{equation} 

\noindent and the cluster size distribution n\( _{s} \)(N) that gives the
probability of having trees of size $s$ per unit area of the substrate 
is given by

\begin{equation}
n_{s}(N)=N_{s}(N)/L^{d}.  \label{Eq.8}
\end{equation} 

Replacing Eqs. (6) and (8) in (7) gives \cite{RV,MM}

\begin{equation}
N=\sum _{s}sn_{s}(N).  \label{Eq.9}
\end{equation} 

During the competition between trees along the evolution of the 
aggregate it may occur that the existence of large neighboring 
trees may inhibit the growing
of smaller ones. This competing process ultimately leads to the death of some
trees that become ''frozen'' within the underlying aggregate. These prevailing
large trees continue the competition within more distant trees in a dynamic
process. Since this situation takes place on all scales, it is reasonable to
expect that the cluster size distribution should exhibit a power-law behavior
so that

\begin{equation}
n_{s}(N)\sim s^{-\tau }f(s^{\sigma }/N),  \label{Eq.10}
\end{equation} 

\noindent where \( \tau  \) is an exponent and \( f(y) \) is a scaling (cut
off) function so that their asymptotic behavior is 
given by \( f(y)\approx 1 \)
for $ y \ll 1$ and $f(y)\approx 0$ for $y \gg 1$.

Substituting (10) in (9) and after some algebra one can obtain \cite{RV} 

\begin{equation}
\sigma =2-\tau.  \label{Eq.11}
\end{equation} 

Furthermore, recalling that the competition among trees actually is a dynamic
process evolving in time (\( t \propto N_{t} \)), one has that the survival
time distribution of the trees (n\( _{t} \)) may also obey a simple power-law
behavior given by

\begin{equation}
n_{t}\propto t^{-\gamma },  \label{Eq.12} 
\end{equation}

\noindent where \( \gamma  \) is an exponent. Therefore, one can establish a
relationship between the size and the survival time \( t \) of the dead trees
so that 

\begin{equation}
 s\sim t^{1/x},  \label{Eq.13} 
\end{equation} 

\noindent where the exponent \( x \) is given by 

\begin{equation}
x=(\tau -1)/(\gamma -1).  \label{Eq.14} 
\end{equation}

Also, it is reasonable to expect that for the trees \(h_{s}\propto t\), so
inserting this relation in equation (13) and comparing it with equation (4)
gives  
\begin{equation}
x=\nu _{\parallel }.  \label{Eq.15} 
\end{equation}

Let us now establish a relationship between the growing trees 
and the properties
of the aggregate. Assuming that the aggregate is self-similar (even compact)
with fractal dimension D, its rms height $h$ is expected
to scale as

\begin{equation}
 h\sim N^{\nu _{l}},  \label{Eq.16} 
\end{equation}

\noindent for \( h<<L \) , where \( \nu _{l} \) is an exponent \cite{MM}.

However, for length scales \( >>h \) the aggregate is uniform in the remaining
\( d \) lateral directions. This fact, in connection to equations (6) and
(16) gives \cite{MM}

\begin{equation}
 D=\nu _{l}^{-1}+d.  \label{Eq.17} 
\end{equation}

The rms height $h$ of the aggregate can be rewritten as

\begin{equation}
 h^{2}=\frac{1}{N_{T}}\sum _{i}h_{i}^{2}=\frac{1}{NL^{d}}\sum _{s}(\sum
_{i\in s}h_{i}^{2})N_{s}(N),  \label{Eq.18} 
\end{equation}

\noindent where \( h_{i} \) is the distance of the i-th particle from the
substrate. Taking into account that the average rms height of a tree of size s
is given by

\begin{equation}
 h_{s}^{2}=\frac{1}{s}\sum _{i\in s}h_{i}^{2},  \label{Eq.19} 
\end{equation}

\noindent equation (18) can be rewritten as

\begin{equation}
 h^{2}=\frac{1}{N}\sum _{s}sh_{s}^{2}n_{s}(N),  \label{Eq.20} 
\end{equation}

\noindent where equation (8) has been used.

Replacing equations (4),(10) and (16) in (20) and performing some algebra one
obtains the scaling relationship \cite{MM}

\begin{equation}
2\nu _{l}=-1+(2+2\nu _{\parallel }-\tau )/\sigma.  \label{Eq.21} 
\end{equation}

Now, using equation (11) and (17) it follows that

\begin{equation}
\sigma =\nu _{\parallel }/\nu _{l}=\nu _{\parallel }(D-d),  \label{Eq.22} 
\end{equation}

\noindent and substituting equation (22) in equation (11) one has \cite{MM}

\begin{equation}
 \tau =2-\nu _{\parallel }(D-d).  \label{Eq.23} 
\end{equation}

The scaling relationship (23) links the exponent \( \tau  \) related to the
tree size distribution to the correlation length exponent of the aggregate
\( \nu _{\parallel }. \) It is interesting to notice that \( \nu _{\perp } \)
is absent in equation (23) due to the fact that during the competitive growth
process larger trees prevent smaller ones from growing. Scale invariance in the
size distribution of the trees (equation(10)) is a result of this competitive
process that is mainly governed by the tree hight (equation(4)).

A typical example for the application of equation (23) is the growth 
of compact aggregates where \(D=d+1\). So, equation (23) becomes

\begin{equation}
 \tau =2-\nu _{\parallel }.  \label{Eq.24} 
\end{equation}

Also, using equations (14), (15) and (23) one obtains

\begin{equation}
\gamma =1/\nu _{\parallel }.  \label{Eq.25} 
\end{equation}

There is no  reason to expect that a 
compact aggregate may result from the addition of compact trees. 
In fact, the volume of trees of size \( s \) (\( v_{s} \)) scales as

\begin{equation}
 v_{s}\sim h_{s}w_{s}^{D-1}\sim s^{\nu _{\parallel }+(D-1)\nu _{\perp }}, 
\label{Eq.26}
\end{equation}

\noindent and we define the following relationship between the volume and the
number of particles \( v\sim s^{\pi } \), 
where \( \pi  \) is an exponent, so one has

\begin{equation}
 \nu _{\parallel }+(D-1)\nu _{\perp }=\pi, 
\label{Eq.27}
\end{equation} 

\noindent where for compact trees one has \( \pi =1 \) while for
non-compact trees one has $\pi > 1$.

Identifying the parallel correlation length with \( w(s) \) and the time with
\( h(s) \), one has \cite{MMM} 

\begin{equation}
z=\nu _{\parallel }/\nu _{\perp }. 
\label{Eq.28}
\end{equation} 

Using equations (14), (15), (23), (27) and (28) for trees
in a compact aggregate (\( D = d+1\)), we can obtain the following
relationships

\begin{equation}
\nu _{\parallel }=\pi z/(z+D-1),
\label{Eq.29}
\end{equation}
 
\begin{equation}
\nu _{\perp }=\pi /(z+D-1), 
\label{Eq.30}
\end{equation}
 
\begin{equation}
\tau =((2-\pi )z+2(D-1))/(z+D-1), 
\label{Eq.31}
\end{equation}

\noindent and
 
\begin{equation}
\gamma =(z+D-1)/(\pi z) 
\label{Eq.32}
\end{equation}

All these relationships establish links among exponents that characterize the
self-affine nature of the interface of the aggregate ($z=\alpha /\beta $) and
those corresponding to the description of the treeing process occurring in the
bulk ($\nu _{\parallel }$,$\nu _{\perp }$,$\tau $,$\gamma$,$\pi$).\\

{\bf III. The RDSR and BD models. Simulation
methods and mesoscopic equations.}\\

In the random deposition with surface relaxation (RDSR) model a particle is
released from a random position above the surface and falls vertically until it
reaches the top of the selected column. Of course, such particle is initially
located at a  distance larger than the maximum height of the interface. The
deposited particle is allowed to relax to a nearest 
neighbor column if the height of the neighboring column is lower than 
the one corresponding to the selected column. Further details on RDSR
aggregates can be found in references \cite{5,edwi}.  
Concerning the dynamics of tree formation, it is worth remembering the
procedure used to define a tree. In the case of the RDSR model,
a newly deposited particle is assumed to belong to the tree 
corresponding to the impingement site, before its eventual relaxation.

The ballistic deposition (BD) model is rather 
simple to describe: a particle is released from a random position above the 
interface of length $L$. Subsequently, the 
particle follows a straight vertical trajectory until it reaches the 
interface, whereupon it sticks. In contrast to the RDSR model, in the BD model
no further relaxation of the particle is considered. Snapshot configurations of
BD  aggregates, and further details on the deposition rules can be found in 
references \cite{5,KPZ}.  Let us also remember that, in this case, 
trees are formed 
assuming that any newly deposited particle belongs to the same
tree as that of the nearest neighbor particle where it is attached. If the
deposited particle has more than one nearest neighbor belonging to different
trees, one of them is selected at random and the particle is incorporated into
that tree.

RDSR and BD aggregates are grown in the direction perpendicular 
to $d$-dimensional substrate, i.e. in $(d + 1)$-dimensions, using samples
of  different sizes ($L^{d}$) with $1 \leq d \leq 5$. Simulation
results are averaged over many different runs, depending on $L$ and $d$. A
Monte Carlo time step (mcs) involves the deposition of $L^{d}$ particles.  
 
The interface of these aggregates is defined as the set of particles  
that are placed at the highest position of each column.  
So, the mean height of the interface (equation(1)) and the interface width
(equation(2)) can be calculated and both are measured in lattice
 units (LU). It is well known that these aggregates are self-affine 
and the width of the 
growing interface $w(L,t)$ scales as equation (3) \cite{5}. 

In contrast to the microscopic details of the growing mechanisms 
of the interface in both models, continuous equations focus on the  
macroscopic aspects of the roughness. Essentially, the aim is to  
follow the evolution of the coarse-grained height function $h(x,t)$ 
using a well-established phenomenological approach that takes into account 
all the relevant processes that survive at a coarse-grained level. 

This
procedure normally leads to stochastic nonlinear partial differential
equations that may be written as follows \cite{5,edwi,KPZ,kar,kar1,kar2}

\begin{equation}
\frac{\partial h({\bf x},t)}{\partial t} = G_{i}\{h({\bf x},t)\} +  F + \eta ({\bf x},t),   \label{eq14} 
\end{equation} 

\noindent where the index $i$ symbolically denotes different 
processes, $G_{i}\{h({\bf x},t)\}$ is a local functional that contains the  
various surface relaxation phenomena and only depends on the  
spatial derivatives of $h({\bf x},t)$ since the growth process 
is determined by the local properties of the surface. 
Also, $F$ denotes the mean 
deposition rate and $\eta ({\bf x},t)$ is the deposition 
noise that determines  
the fluctuations of the incoming flux around its mean value $F$.    
It is usually assumed that the noise is spatially and temporally 
uncorrelated, so fluctuations are given by a Gaussian white noise 
 
\begin{equation} 
\left\langle \eta ({\bf x},t)\right\rangle =0 , 
\label{eq15} 
\end{equation} 

\noindent and 
 
\begin{equation} 
\left\langle \eta ({\bf x},t)\eta ({\bf x`},t`)\right\rangle = 
2C\delta ^{d}({\bf x} - {\bf x`})\delta
(t-t`),  \label{eq16} 
\end{equation} 

\noindent where the brackets denote ensemble averaging, $C$ is 
the strength of the  
fluctuations, and $d$ the spatial dimension of the surface. 

The RDSR model can be described by the 
Edwards-Wilkinson equation given by \cite{5,edwi,kar,55} 
  
\begin{equation}  
\frac{\partial h({\bf x},t)}{\partial t} = 
F + \nu_{o} \nabla ^{2}h({\bf x},t)+\eta ({\bf x},t), 
\label{eq108} 
\end{equation}  

\noindent where $\nu_{o}$ plays the role of an effective surface tension,  
since the $\nu_{o} \nabla^{2}h({\bf x},t)$ term tends to smooth the interface. 
Equation (\ref{eq108}) can be solved in Fourier space and the following 
values of the exponents are obtained: $z = 2$, $\alpha = (2 - d)/2$ 
and $\beta = (2 - d)/4$. So, there exists an upper critical dimension $d_{c}=2$
above which one has $\alpha =0$ and $\beta = 0$. 

The BD model can be described by the 
Kardar-Parisi-Zhang equation given by \cite{KPZ} 
  
\begin{equation}  
 \frac{\partial h({\bf x},t)}{\partial t} = 
F + \nu \nabla ^{2}h+\frac{\lambda}{2}(\nabla h)^{2}+\eta ({\bf x},t) ,
\label{eq18} 
\end{equation}  

\noindent where $\lambda$ plays the role of an effective lateral growth. 
In this equation, the exponents satisfy the scaling relation

\begin{equation}  
z+\alpha=2 ,
\label{eq123} 
\end{equation}

\noindent as a consequence of Galilean invariance in the related Burgers
equation \cite{20}. In the usual field theory approach, a perturbation theory
is defined with respect to the nonlinear term ($\lambda$). The corresponding
renormalization group reveals \cite{21} that the physics of the KPZ is related
to a strong coupling fixed point, which is inaccessible by perturbative
methods. Except for $d=1$, where, thanks to a fluctuation-dissipation theorem,
an analytic solution is possible. In this case one has $\alpha = 1/2$, $\beta
= 1/3$, so that $z = 3/2$. Extensive numerical simulations have been
carried out for restricted solid on solid models \cite{22}. These are discrete
models that belong to the KPZ universality class \cite{23} and these studies
show a gradual decrease in the value of $\alpha$, without any
evidence for an upper
critical dimension. In the same direction a recent study performed using a
nonperturbative renormalization of the KPZ equation suggests a gradual decrease
in the value of $\alpha$ \cite{it}. On the other hand, several analytical
approaches \cite{24} suggest an upper critical dimension $d_{c}=4$ above which
one should have $\alpha =0$.\\

{\bf IV. Results and discussion. }\\ 

Figure 2 shows the tree distribution functions corresponding to the RDRS model
in dimension \( 1\leq d\leq 5 \). The tree
size distribution (figure 2 (a)) for \( d=1 \) exhibits a power law according
to Eq.(10) with exponent  \( \tau =1.35\pm 0.03 \). For
higher dimensions \( n_{s} \) also exhibits power law behavior but the slope
\( \tau =1.50\pm 0.01 \) becomes independent of the dimensionality. A similar
behavior is shown by the survival time distribution (figure 2(b)) that
has slopes \( \gamma =1.54\pm 0.03 \) for \( d=1 \) and 
\( \gamma =2.00\pm 0.03 \) for
\( d\geq 2 \), respectively. The measured exponents are listed in Table I.

The distribution functions of the BD model also exhibit power 
law behavior (figure
(3)). However, in contrast to the case of RDSR, for both \( n_{s} \) (figure
3(a)) and \( n_{t} \) (figure 3(b)) the slopes depend on the dimensionality.
The measured exponents are listed in the $5^{th}$ and $9^{th}$ columns of Table
II.

The behavior of the rms height of the trees as a function of the tree size \( s
\) is shown in figures 4(a) and 4(b) for RDSR and BD, respectively. Again, nice
power laws are observed in all cases. For the RDSR model a clear change in the
slope is observed between \( d=1 \) and \( d\geq 2 \), where for the higher
dimensions the slopes are independent of the dimensionality (see $3^{rd}$
column of Table I). For the BD model the slopes are different in all dimensions
(see $3^{rd}$ column of Table II).

Results shown in figure 5 for the rms width of the trees (\( w_{s} \)) versus
\( s \) are also fully consistent with the previous results and are listed in
the $2^{nd}$ column of Tables I and II. 

Based on our numerical estimation of the exponents \( \tau  \), \( \gamma  \),
\( \nu _{\perp } \) and \( \nu _{\parallel } \) we are in a condition
to check the scaling relationships discussed in section II that are summarized
by equations (24), (27), (28), (31), (32) and (38), where the latter only holds
for the BD model.

The results obtained for the RDSR model (Table I) show that 
the exponents \( \nu _{\perp } \),
\( \nu _{\parallel } \), \( \tau  \) and \( \gamma  \) are independent of
the dimensionality of the surface for \( d\geq 2 \), as expected from figures
1-4. These results are consistent with the fact that $d = 2$ is the 
upper critical dimension of the EW universality class.

Inserting the exact value \( z=2 \) for the EW universality class in equation
(31) the "theoretical" estimation of \( \tau  \) can be obtained (\( 7^{th} \)
column of Table I), which is in excellent agreement with the numerical data.
Also, inserting the measured values of \( \nu _{\parallel } \) in equation
(24) (see the \( 6^{th} \) column of Table I) excellent agreement with the
measured exponents is obtained. The quality of the achieved agreement is also
excellent when comparing theoretical and estimated (equation (25)) values of \(
\gamma  \), which is shown in the \( 10^{th} \) and \( 9^{th} \) columns
of Table I, respectively.

For the case of the RDSR model it is interesting to notice that the exponent
\( \pi  \) (\( 4^{th} \) column of Table I) departs from the trend
exhibited by the other exponents. In fact, while \( \pi \sim 1 \) for \( d\leq
2 \) , such an exponent increases monotonically with the dimensionality for \(
d>2 \). Roughly, an average increment of \( 1/4 \) in \(
\pi  \) for each additional dimension is estimated. The value of \( \pi =1 \)
corresponds to the development of compact trees (see equations (26),(27)), so
it is concluded that for \( d>2 \) the trees leading to the formation of RDSR
aggregates are non-compact objects. Therefore, the branches of the trees
become interweaved forming complex patterns.

Our results for the BD model are listed in Table II. In this case the measured
exponents (\( \nu _{\perp } \), \( \nu _{\parallel } \), \( \tau  \) and
\( \gamma  \) ) depend on the dimensionality, as expected from 
figures \( 2-5 \).
The measured values of \( \tau  \) (\( 5^{th} \) column of Table II) are in
agreement with the estimation obtained using equation (24) (\( 6^{th} \) column
of Table II) for \( d\leq 4 \). For \( d=5 \) this estimation presents a small
deviation due to the large error involved in the evaluation of the exponent.
However, a slight increment of \( \tau  \) with the dimensionality is found in
both cases (the measured exponent and equation (24)). This result
is in contrast to the conjecture stating that $d =4$ may be the upper 
critical dimension of the KPZ universality class \cite{22}. Furthermore, using
equations (31), (38) and the values of \( \alpha  \) reported in reference
{[}25{]} two additional comparisons can be made: inserting the 
values of \( z \)
obtained using renormalization group calculations (numerical simulation)
in equation (31) gives the values of \( \tau  \) listed in the \( 7^{th} \)
(\( 8^{th} \)) column of Table II. In both cases, excellent agreement with
the measured exponents is found.

We have to recognize large errors in our evaluation of \( \gamma  \) for the
BD model (see figure 2(b)) that prevent reliable estimations for \(
d\geq 4 \). However, comparisons with results obtained using equation (25) (\(
10^{th} \)column of Table II), renormalization group calculation (\( 11^{th}
\) column of Table II) and numerical evaluation (\( 12^{th} \) column of Table
II), exhibit the trend already observed by our direct measurement of \( \gamma 
\).

It should also be noticed that \( \pi \sim 1 \) (within error bars) pointing
out that BD aggregates are built up by compact trees at least for \( d\leq 4
\), in contrast with our previous results obtained for RDSR aggregates for
$d>2$.\\

{\bf  CONCLUSIONS.}\\

Extensive numerical evidence is provided showing that the bulk of two
typical growth models, belonging to both the KPZ and EW universality classes,
can be rationalized in terms of a treeing process. Scaling relationships,
linking exponents relevant to the treeing process with standard dynamic
exponents are reviewed and tested numerically in dimensions \( 1\leq
d \leq 5 \). So, our main conclusions are summarized in Tables I and II where
all the measured exponents are listed and compared to theoretical
predictions. It is concluded that BD trees are compact objects
(at least up to $d=4$) while RDSR trees are non-compact objects 
for $d>2$ leading to
interweaved complex structures. Based on
the validity of equations (29-32) and 
since above the upper critical dimension the exponents $\nu_{\parallel }$,
$\nu _{\perp }$, $\tau $, and  $\gamma$ remain constant for
self-similar aggregates, it can be stated that above the upper critical
dimension the change in the dimensionality of the aggregate implies that $\pi$
must also change. Therefore, we conjectured that all kinds of self-affine 
aggregates above their upper critical dimension can be rationalized in 
term of the superposition of non-compact trees. 

Summing up, it is shown that the evolution of the treeing
statistics is a suitable method for the characterization of the growing
aggregates with promising applications in experimental situations.
In fact, it is well known that the electrochemical deposition method 
in thin cells provides a rather simple experimental setup for the 
observation of a wide variety of self-affine growing patterns \cite{ec1,ec2}.
In this case the dynamics of growing trees leading to frozen structures
can easily be observed and the evaluation of the tree size distribution
can  be obtained straightforwardly after proper digitalization of the
images of the aggregates. Another field of application is the 
study of polycrystalline thin film growth by scanning tunneling 
microscopy \cite{stm1,stm2}
and atomic force microscopy \cite{afm}.
In these cases the direct evaluation of the number of crystallites
as a function of the deposit height accounts for the number
of frozen crystallites (i.e. the trees of the growing aggregate)
and the crystallite size distribution can be evaluated.\\  

{\bf  ACKNOWLEDGMENTS}. This work was financially supported by 
CONICET, UNLP and ANPCyT (Argentina).

\newpage

{\bf Table I:} List of exponents \( \nu _{\perp } \), \( \nu _{\parallel }
\), \( \tau  \) (\( 5^{th} \) column) and \( \gamma  \) (\( 8^{th} \) column)
measured for the RDSR model for dimensions \( 1\leq d\leq 5 \). The values of
theoretical estimations of \( \tau  \) and \( \gamma  \) are obtained taking
\( z=2 \) in Equation (31) (\( 7^{th} \) column) and Equation (32) (\( 10^{th}
\) column), respectively. For these estimations \( \pi =1 \) for \( d\leq 2 \)
and \( \pi =5/4, \) \( 3/2, \) \( 7/4 \) for \( d=3,4,5 \), have been used,
respectively. The values of \( \pi  \) obtained using Equation (27) are shown
in the \( 4^{th} \) column. The values of \( \tau  \) and \( \gamma  \)
obtained using Equations (24) and (25) are shown in the \( 6^{th} \) and
\( 9^{th} \) columns, respectively.\\ 

\vspace{0.3cm}
{\centering \begin{tabular}{|c|c|c|c|c|c|c|c|c|c|}
\hline 
\( D \)&
\( \nu _{\perp } \)&
\( \nu _{\parallel } \)&
\( \pi  \)&
\( \tau  \)&
\( \tau  \)(Eq.24)&
\( \tau  \)(t)&
\( \gamma  \)&
\( \gamma  \)(Eq.25)&
\( \gamma  \)(t)\\
\hline 
\hline 
1+1&
0.35(2)&
0.63(1)&
0.98(2)&
1.35(3)&
1.37(1)&
4/3&
1.54(3)&
1.59(3)&
3/2\\
\hline 
2+1&
0.27(2)&
0.50(1)&
1.02(4)&
1.49(1)&
1.50(1)&
3/2&
1.98(5)&
2.00(4)&
2\\
\hline 
3+1&
0.26(2)&
0.50(1)&
1.28(6)&
1.50(3)&
1.50(1)&
3/2&
2.00(2)&
2.00(4)&
2\\
\hline 
4+1&
0.25(2)&
0.50(1)&
1.50(8)&
1.50(1)&
1.50(1)&
3/2&
1.99(2)&
2.00(4)&
2\\
\hline 
5+1&
0.25(1)&
0.49(1)&
1.74(4)&
1.50(2)&
1.51(1)&
3/2&
1.99(4)&
2.04(4)&
2\\
\hline 
\end{tabular}\par}
\vspace{0.3cm}

\newpage

{\bf Table II:} List of exponents \( \nu _{\perp } \), \( \nu _{\parallel }
\), \( \tau  \) (\( 5^{th} \) column) and \( \gamma  \) (\( 9^{th} \) column)
measured for the BD model for dimensions \( 1\leq d\leq 5 \). The estimations
performed using Equations (24) and (25) are shown in the \( 6^{th} \) and \(
10^{th} \) columns, respectively. The exponents obtained using Equations (31),
(32) and (38), \( \pi =1 \) and \(
\alpha  \) values obtained by renormalization group calculation (RG) and
numerical simulations (NS) reported in reference \cite{it} are shown in the \(
7^{th} \), \( 8^{th} \), \( 11^{th} \), \( 12^{th} \) columns. The exponent \(
\pi  \) is evaluated using equation (27) (\( 4^{th} \) column).\\
\vspace{0.3cm} {\centering \begin{tabular}{|c|c|c|c|c|c|c|c|c|c|c|c|} \hline 
\( D \)& \( \nu _{\perp } \)&
\( \nu _{\parallel } \)&
\( \pi  \)&
\( \tau  \)&
\( \tau  \)(Eq.24)&
\( \tau  \)(RG)&
\( \tau  \)(NS)&
\( \gamma  \)&
\( \gamma  \)(Eq.25)&
\( \gamma  \)(RG)&
\( \gamma  \)(NS)\\
\hline 
\hline 
1+1&
0.40(1)&
0.60(1)&
1.00(2)&
1.40(1)&
1.40(1)&
1.40&
1.40&
1.70(2)&
1.66(3)&
1.66&
1.66\\
\hline 
2+1&
0.29(1)&
0.45(1)&
1.03(2)&
1.57(1)&
1.55(1)&
1.54&
1.55&
2.3(1)&
2.22(5)&
2.22&
2.24\\
\hline 
3+1&
0.23(2)&
0.38(1)&
1.07(6)&
1.65(1)&
1.62(1)&
1.63&
1.64&
3.2(2)&
2.63(7)&
2.75&
2.77\\
\hline 
4+1&
0.21(4)&
0.34(2)&
1.18(16)&
1.70(2)&
1.66(2)&
1.69&
1.70&
-&
2.94(15)&
3.27&
3.30\\
\hline 
5+1&
-&
0.33(4)&
-&
1.75(3)&
1.67(4)&
1.73&
1.73&
-&
3.0(3)&
3.78&
3.76\\
\hline 
\end{tabular}\par}

\newpage

\begin{figure} 
\narrowtext 
\centerline{{\epsfysize=2.0 in \epsffile{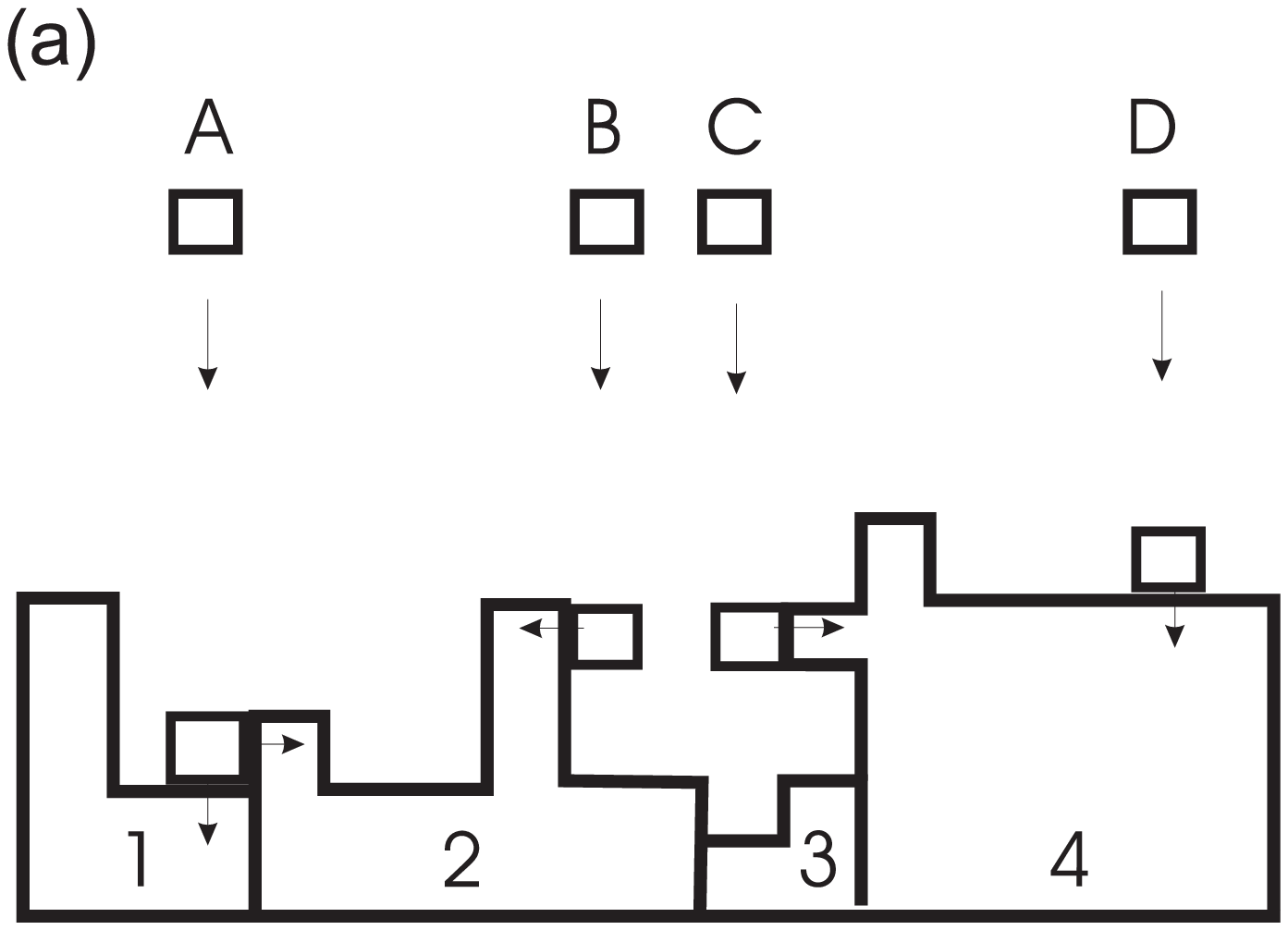}}} 
\centerline{{\epsfysize=2.0 in \epsffile{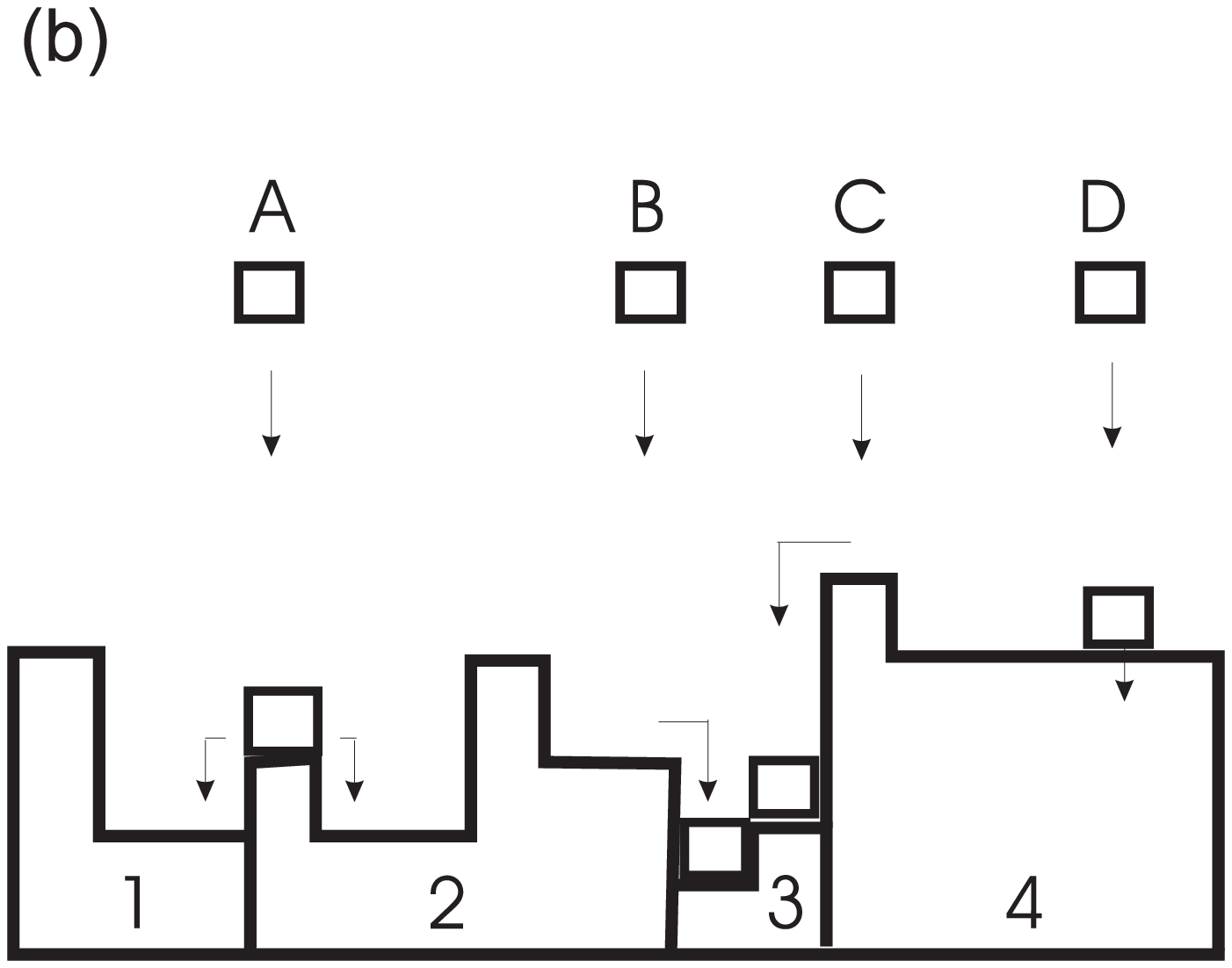}}}
\caption{ Schematic view of the deposition of particles forming trees. In
(a) the particles are deposited with the rules of ballistic deposition.
Particle A belongs to tree 1 or 2 with the same probability, particle B
belongs to tree 2 and particles C and D belong to tree 4. The growth of tree
3 is stopped after particle C deposition and it becomes a frozen tree. In (b)
the particles are deposited with the rules of random deposition with surface
relaxation. Particle A, which belongs to tree 2, can either relax to the
right or to the left with the same probability. Notice that in the former, 
particle A partially shadows tree 1, while in the latter it simply becomes
attached to tree 2. Particle B belongs to tree 2 but it relaxes on top of tree
3 causing partial shadowing. Particle C belongs to tree 4 and after relaxation
on top of tree 3 causes the growth of that tree to stop, so that it becomes a 
frozen tree. Finally, particle D belongs to tree 4.  } 
\label{fig1} 
\end{figure}

\begin{figure} 
\narrowtext 
\centerline{{\epsfysize=4.0 in \epsffile{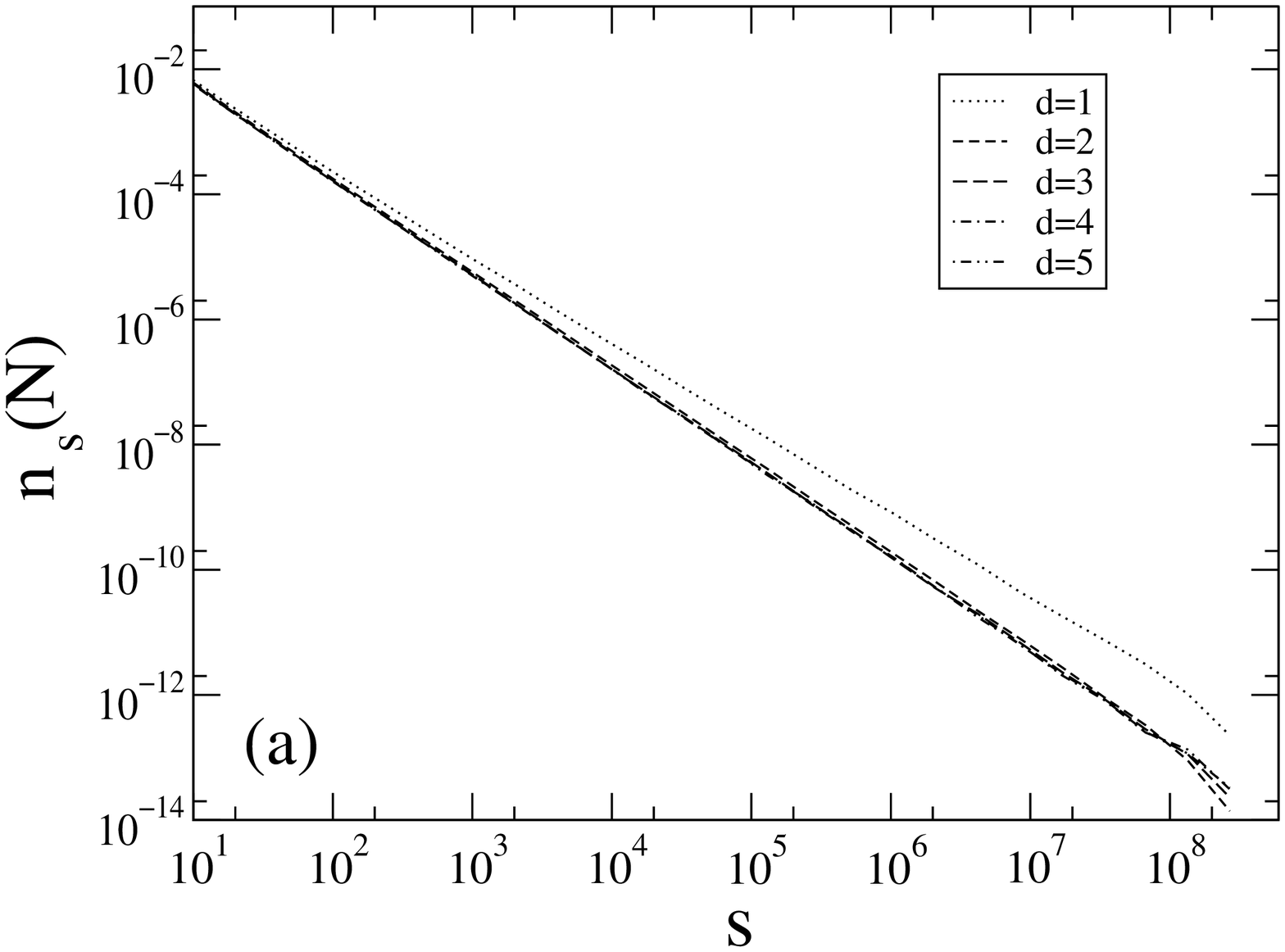}}} 
\centerline{{\epsfysize=4.0 in \epsffile{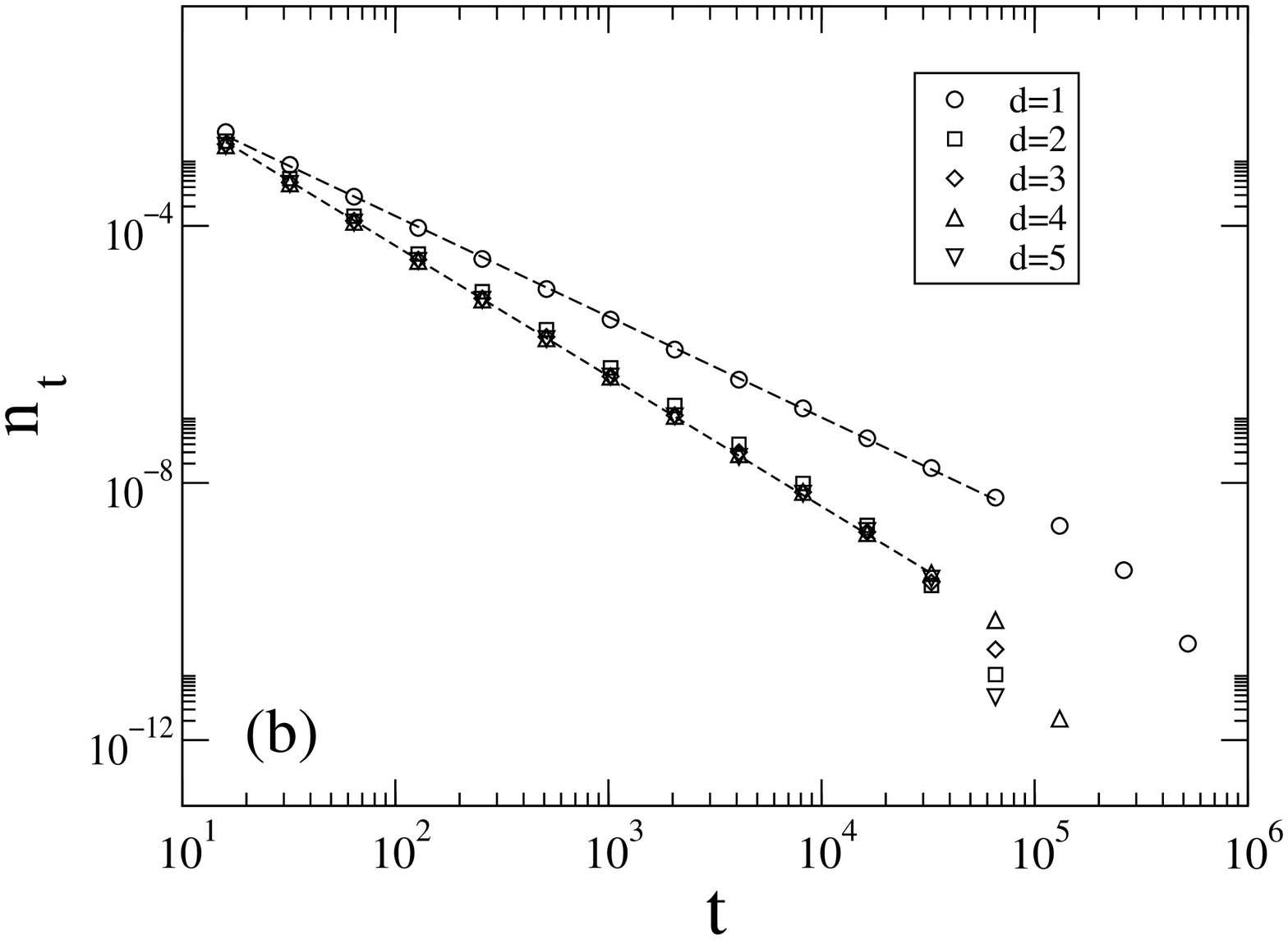}}} 
\caption{ Log-log plots of the tree distribution functions corresponding to
the RDRS model in dimensions \( 1\leq d\leq 5 \). (a) The tree size
distribution, where the size $s$ is given by the number of particles of 
each tree. (b) The
survival time distribution, where the time is measured in mcs. } 
\label{fig2} 
\end{figure}

\begin{figure} 
\narrowtext 
\centerline{{\epsfysize=4.0 in \epsffile{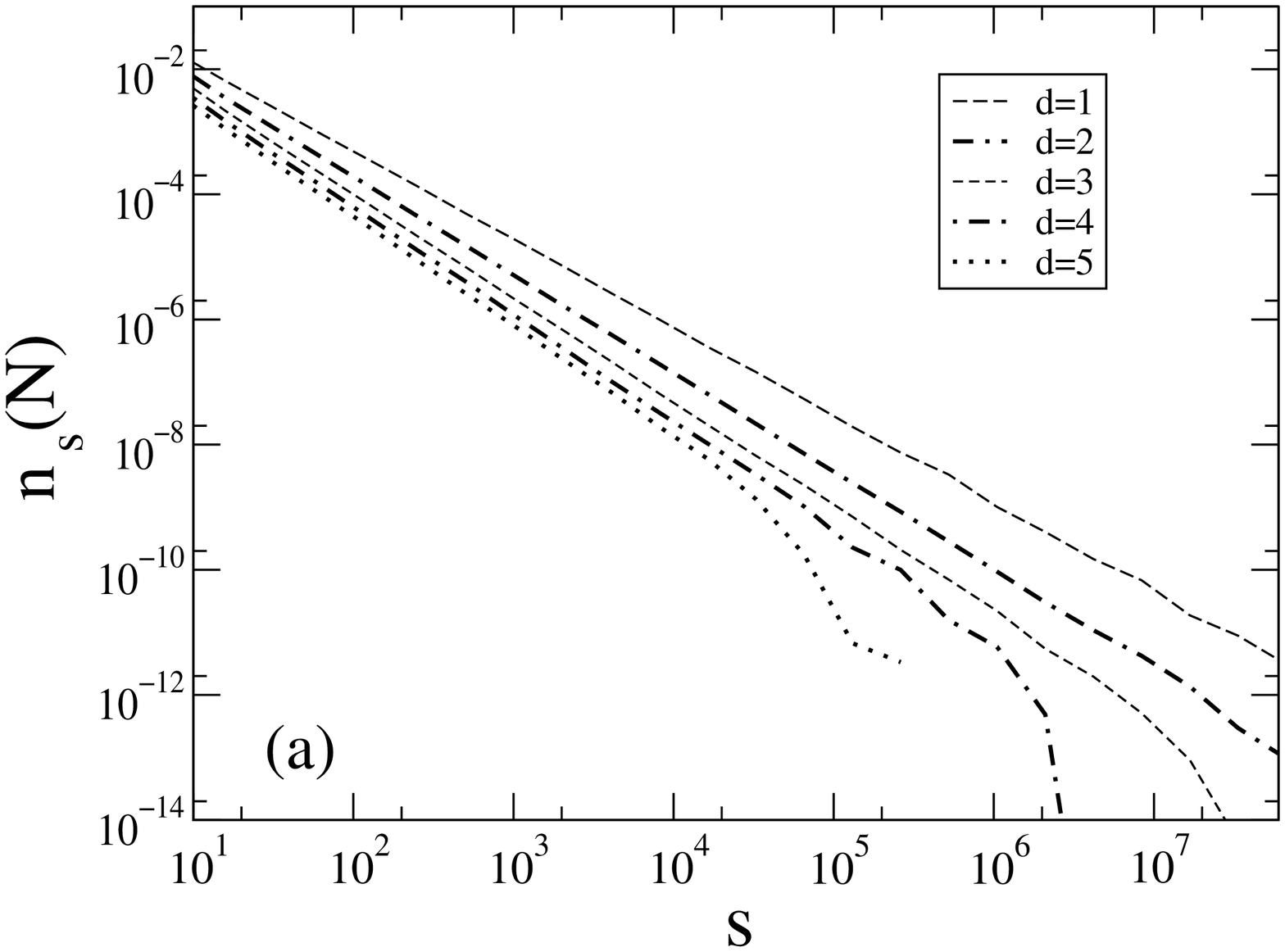}}} 
\centerline{{\epsfysize=4.0 in \epsffile{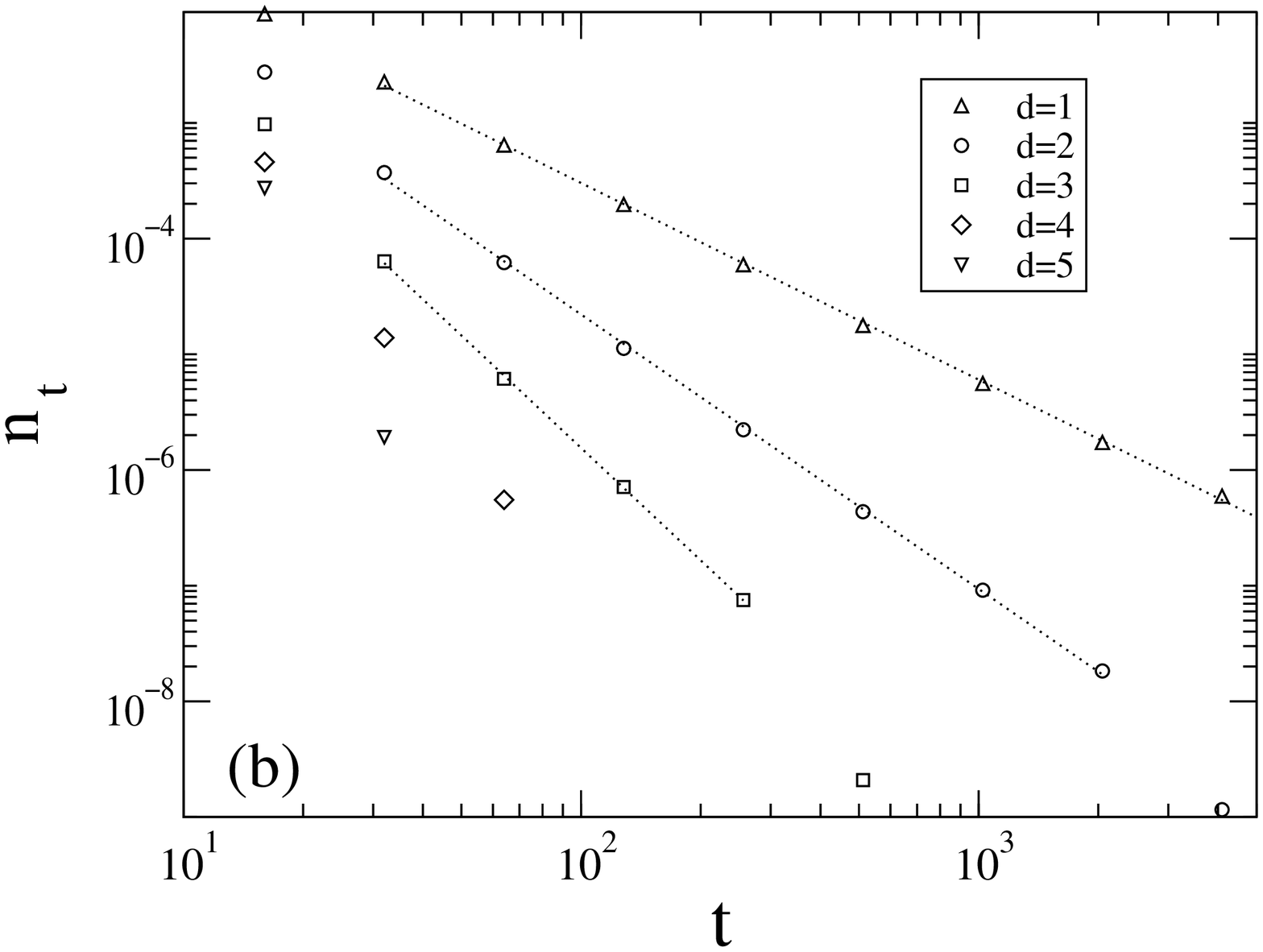}}} 
\caption{ Log-log plots of the tree distribution functions corresponding to the
BD model in dimensions \( 1\leq d\leq 5 \). (a) The tree size distribution
where the size $s$ is given by the number of particles of 
each tree. (b) The survival time distribution, where the time is measured in mcs. } 
\label{fig3} 
\end{figure}

\begin{figure} 
\narrowtext 
\centerline{{\epsfysize=4.0 in \epsffile{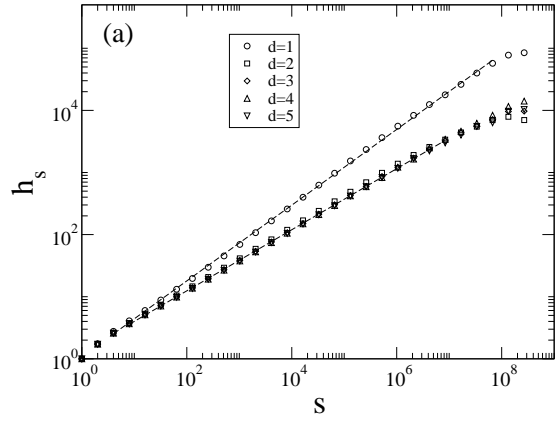}}} 
\centerline{{\epsfysize=4.0 in \epsffile{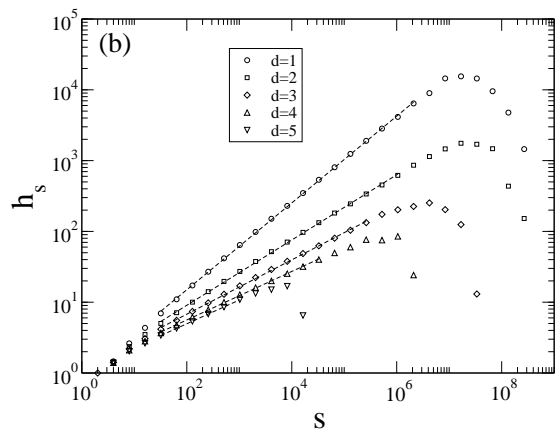}}} 
\caption{ (a) and (b) Log-log plots of rms height of the trees as a function of
the tree size \( s \) for the RDSR model and the BD model, respectively. 
The rms height is measured in LU and the size $s$ is given by the number 
of particles of each tree. } 
\label{fig4} 
\end{figure}

\begin{figure} 
\narrowtext 
\centerline{{\epsfysize=4.0 in \epsffile{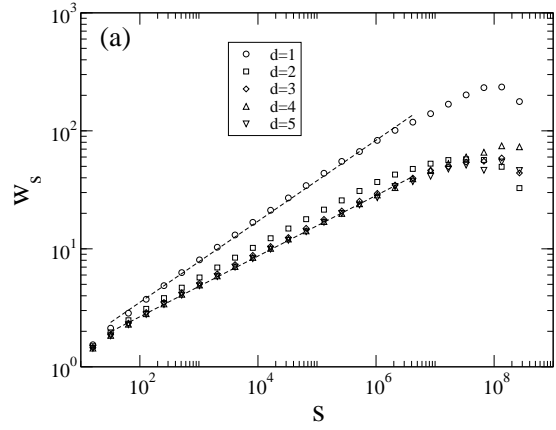}}} 
\centerline{{\epsfysize=4.0 in \epsffile{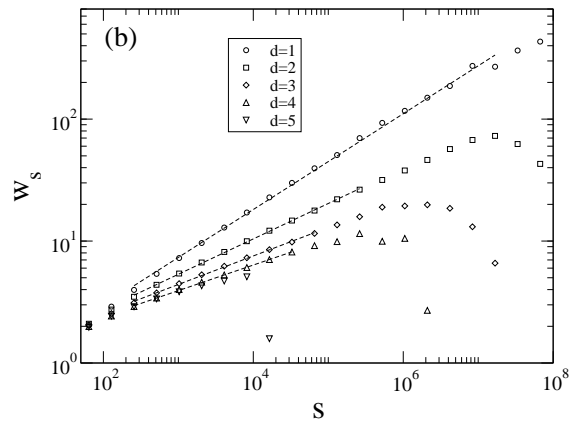}}} 
\caption{ (a) Log-log plots of rms width of the trees as a function of the
tree size \( s \) for the RDSR model. (b) Log-log plots of rms width of the
trees as a function of the tree size \( s \) for the BD model. The rms width 
is measured in LU and  the size $s$ is given by the number 
of particles of each tree. } 
\label{fig5} 
\end{figure}

\end{document}